\documentstyle[12pt]{article}
\newcommand{\be}{\begin{equation}}
\newcommand{\ee}{\end{equation}}
\newcommand{\ben}{\begin{eqnarray}\displaystyle}
\newcommand{\een}{\end{eqnarray}}
\newcommand{\refb}[1]{(\ref{#1})}

\begin{document}

{}~ \hfill\vbox{\hbox{hep-th/9705212}\hbox{MRI-PHY/970512}}\break

\vskip 3.5cm

\centerline{\large \bf Kaluza-Klein Dyons in String Theory}

\vspace*{6.0ex}

\centerline{\large \rm Ashoke Sen\footnote{On leave of absence from 
Tata Institute of Fundamental Research, Homi Bhabha Road, 
Bombay 400005, INDIA}
\footnote{E-mail: sen@mri.ernet.in, sen@theory.tifr.res.in}}

\vspace*{1.5ex}

\centerline{\large \it Mehta Research Institute of Mathematics}
 \centerline{\large \it and Mathematical Physics}

\centerline{\large \it Chhatnag Road, Jhusi, Allahabad 221506, INDIA}

\vspace*{4.5ex}

\centerline {\bf Abstract}

S-duality of hetertotic / type II string theory compactified on a
six dimensional torus requires the existence of Kaluza-Klein
dyons, carrying winding charge. We identify the zero modes of the
Kaluza-Klein monopole solution which are responsible for these
dyonic excitations, and show that we get the correct degeneracy
of dyons as predicted by S-duality. The self-dual harmonic two
form on the Euclidean Taub-NUT space plays a crucial role in this
construction.

\vfill \eject

\baselineskip=18pt

Type II / heterotic string theory compactified on $T^6$ 
has been conjectured to have an exact SL(2,Z) S-duality symmetry
that forms a subgroup of the full duality group of these
theories\cite{ASREV,HT}. Acting on an elementary string state carrying pure
winding charge along one of the compact directions, the S-duality
transformation produces a state carrying Kaluza-Klein magnetic
charge, as well as winding charge. The existence of these states
is a prediction of S-duality symmetry. Furthermore, if S-duality
is an exact symmetry of the theory, then these states must have
degeneracy identical to that of the elementary string state
carrying pure winding charge. In this letter we shall verify this
prediction for a class of Kaluza-Klein dyons, namely those with
one unit of magnetic charge.\footnote{Classical solutions
carrying Kaluza-Klein magnetic charge and winding charge can be
constructed by using the standard trick of duality
rotation\cite{TSW} (see, for example,
\cite{MIRJ}).}

Let us denote the six internal directions labelling $T^6$ by
$x^4, \ldots x^9$. 
The momentum and winding charge quantum numbers along $x^4$
are labelled by a 2 dimensional vector:
\be \label{e1}
\vec \alpha = \pmatrix{p\cr w} \, ,
\ee
where $p$ and $w$ are integers.
Suppose $n_K$ and $n_H$ are integers labelling Kaluza-Klein
monopole charge and $H$-monopole charge associated with the $x^4$
direction respectively. Then we
can define another two dimensional vector $\vec \beta$ through
the relation\cite{ASREV}
\be \label{e2}
L\vec \beta = \pmatrix{n_K\cr n_H}\, ,
\ee
where
\be \label{e3}
L = \pmatrix{0 & 1\cr 1 & 0}\, ,
\ee
is the natural metric on the two dimensional lattice of momentum
and winding charges along $x^4$. 
SL(2,Z) S-duality group acts on $(\vec \alpha, \vec\beta)$
as\cite{ASREV}
\be \label{e4}
\pmatrix{\vec\alpha\cr \vec\beta} \to 
\pmatrix{l & q\cr r & s} \pmatrix{\vec\alpha\cr \vec\beta} 
\, , \ee
where 
\be \label{e5}
l,q,r,s\in Z, \qquad ls-qr=1\, .
\ee
An elementary string state carrying one unit of 
winding charge along $x^4$ has
\be \label{e6}
\vec\alpha=\pmatrix{0 \cr 1}, \qquad
\vec\beta=0.
\ee
Acting on this state,
the SL(2,Z) transformation given in eq.\refb{e4} produces a state with
\be \label{e7}
\vec \alpha= \pmatrix{0\cr l}, \qquad \vec \beta=\pmatrix{0\cr
r}\, ,
\ee
{\it i.e.}
\be \label{e8}
w=l, \quad n_K=r\, .
\ee
Standard argument shows that as a consequence of eq.\refb{e5},
$l$ and $r$ are relatively prime. Thus according to S-duality,
for every pair of integers $l$ and $r$ relatively prime, 
the theory must contain dyons with $r$ units of Kaluza-Klein
magnetic charge and $l$ units of winding charge. The degeneracy
of these states must match the degeneracy of the elementary
string states carrying quantum numbers \refb{e6}.

We shall focus on states with $r=1$ and identify the dyonic
excitations of the monopole carrying winding 
charge.\footnote{These are
$T$-dual to the $H$-dyon states analysed in ref.\cite{BLUM}.} 
We begin by
writing down the Kaluza-Klein monopole solution in ten
dimensional string metric\cite{SOR,GP}
\be \label{e9}
ds^2 = - dt^2 + \sum_{m=5}^9 dx^m dx^m + ds_{TN}^2 \, ,
\ee
where $ds_{TN}$ denotes the Euclidean Taub-NUT metric:
\be \label{e10}
ds_{TN}^2 = V\{d x^4 + 4m (1-cos\theta)d\phi\}^2 + V^{-1} (dr^2 + r^2
d\theta^2 + r^2 \sin^2\theta d\phi^2)\, ,
\ee
\be \label{e11}
V = (1 +{4m\over r})^{-1}\, .
\ee
In order for the solution to be non-singular at the origin, we
need the periodicity of $x^4$ to be $16\pi m$.

This solution has three bosonic zero modes, which are simply the
translational modes of the solution in the Euclidean space
labelled by the polar coordinates $(r,\theta, \phi)$. The momenta
conjugate to these zero modes represent spatial momenta of the
monopole. Note that since the solution is invariant under a
translation along the $x^4,\ldots x^9$ direction, there is no
zero mode associated with translation in these directions,
and hence the monopole cannot carry momentum along these
directions. 

If these were all the bosonic deformations of the
solution, there would not be any dyonic excitation of the
solution carrying winding charge along the $x^4$ direction.
However, the presence of the anti-symmetric tensor gauge field
$B_{\mu\nu}$ in this theory, and the existence of a harmonic two
form in the Euclidean Taub-NUT space\cite{BRPO,GALO,WEIN,CONNELL},
allows us to construct new bosonic zero modes of the solution,
by considering deformations of the form:\footnote{For type II
theories, one can get other deformations involving Ramond-Ramond
gauge fields. These will be responsible for constructing dyonic
states carrying Ramond-Ramond charge, as predicted by U-duality,
but we shall not consider them here. Also, one needs to take into account 
these deformations to get the correct counting of zero modes for Kaluza-Klein
monopoles, as given in \cite{HULLNEW}.}
\be \label{e14}
B = \Theta\Omega\, ,
\ee
where $\Theta$ is the deformation parameter, and $\Omega$ is the
self-dual harmonic two form:
\be \label{e12}
\Omega = C{r\over r+4m} (d\sigma_3 + 
{4m\over r(r+4m)} dr\wedge\sigma_3)\, .
\ee
Here $C$ is a normalization constant, and,
\be \label{e13}
\sigma_3 = (4m)^{-1} ( dx^4 + 4m(1-\cos\theta) d\phi)\, .
\ee
The normalization constant $C$ is chosen in such a way that the
imaginary part of the
string action for an Euclidean world sheet, wrapped on a
(non-compact) two cycle dual to $\Omega$, is given by $i\Theta$.
This will make the coordinate $\Theta$ periodic with period
$2\pi$.

Note that $\Omega$ can be written as,
\be \label{e15}
\Omega = d\xi \, ,
\ee
where
\be \label{e16}
\xi = C {r\over r+4m} \sigma_3\, .
\ee
$\xi$ represents a non-singular one form at $r=0$, but does not vanish as
$r\to \infty$. This shows that the zero mode deformation that we
have introduced corresponds to a pure gauge deformation of the
solution, with the gauge transformation parameter not vanishing
at $\infty$. The situation is exactly analogous to the case of
BPS monopoles, for which the collective coordinate conjugate to
the electric charge corresponds to a pure gauge deformation. Since
$\Theta\to \Theta +\alpha$ is a gauge transformation, the
collective Hamiltonian does not depend explicitly on $\Theta$.
Hence the momentum $p_\Theta$ conjugate to $\Theta$ is
conserved, and can be interpreted as the winding 
charge.\footnote{ To see this let us consider a configuration of
constant $p_\Theta$, for which $\Theta=\alpha t +\Theta_0$. Using
eqs.\refb{e14}-\refb{e13} we see that for this solution, as $r\to
\infty$,
$$ H=dB \simeq \alpha C r^{-2} dt\wedge dr\wedge (dx^4 +
4m(1-cos\theta)d\phi) + \alpha C \sin\theta dt \wedge d\theta\wedge
d\phi $$
The term in $H$ proportional to $dt\wedge dr\wedge dx^4$
indicates that the solution carries winding charge
$\propto\alpha\propto p_\Theta$. The other terms can be
interpreted as non-trivial axion background in the four
dimensional theory.}
Upon quantization $p_\Theta$ is quantized in integer units, 
and would correspond to the
winding number $w$.

This establishes the existence of the required dyonic excitations
carrying $n_K=1$, $w$ arbitrary. In order to verify the
predictions of S-duality, we also need to make sure that these
states have degeneracy identical to that of a singly wound 
elementary BPS excitation in the corresponding string theory. For
type II string theory, the degeneracy of the elementary string
state is 256, corresponding to an
ultra-short multiplet of the supersymmetry algebra. For the
Kaluza-Klein dyons, this degeneracy comes from quantizing the
fermionic zero modes associated with the broken supersymmetry
generators. Since the monopole solution breaks half of the thirty
two supersymmetry generators, the sixteen broken generators give
rise to 16 fermionic zero modes, whose quantization gives a
$2^8=256$-fold degenerate state. Thus for type II theory on
$T^6$, we get exactly the right degeneracy for the Kaluza-Klein
dyons.

The situation in the heterotic string theory is somewhat more
complicated. In this case, in the elementary string spectrum, a
BPS state carrying pure winding charge is $16\times 24$ fold
degenerate, corresponding to 24 short multiplets. (The number 24
is related to the number of bosonic oscillators in the
left-moving sector of the theory.) On the other hand, the
Kaluza-Klein monopole solution breaks eight of the sixteen
supersymmetry generators. This gives 8 fermionic zero modes, and
hence $2^4=16$-fold degeneracy, corresponding to one short
multiplet. Thus we are missing a factor of 24. 

The resolution to this problem comes from the fact that the
Kaluza-Klein monopole
solution that we have displayed carries one unit of gravitational
instanton number\cite{GIBHAW}, and hence acts as a source of $-1$
unit of $H$-magnetic charge. These states are dual to elementary
string states with $(p=-1, w=1)$, which indeed have degeneracy
16, corresponding to a single short multiplet.
If we want a state whose magnetic charge
is only of the Kaluza-Klein type, we must cancel this 
$H$-magnetic charge. This can be done by placing a gauge
instanton inside the Euclidean Taub-NUT space. At a generic point
in the moduli space of the theory, the gauge group is 
purely abelian, and hence
the instanton must necessarily have zero size. This corresponds
to a heterotic five-brane wrapped on the $T^5$ labelled by $x^5,
\ldots x^9$. This will have extra fermionic and bosonic zero
modes which must be taken into account in computing the
degeneracy of the state. However, this is precisely the problem
that has been addressed in the context of calculating degeneracy
of H-monopoles\cite{BAN,GH}, and it is known that
the quantization of these degrees of freedom enhances the
degeneracy by a factor of 
24\cite{WITTSM,PORR,SETH}.  In the present case we also get a set
of extra bosonic and fermionic zero modes associated with the location
of the small instanton in the Taub-NUT space and their fermionic 
partners. Quantization of these zero modes gives an extra multiplicative
factor in the degeneracy equal to the number of normalizable
harmonic forms on the Taub-NUT space.
This number is known to be unity\cite{WEIN}.
This then gives the right counting
of the degeneracy of states of Kaluza-Klein dyons for the
heterotic string theory.

Acknowledgement: I wish to thank C. Hull and E. Witten for useful
correspondence.

\end{document}